\documentclass[aps,floatfix,12pt]{revtex4}
\usepackage{graphicx}
\usepackage[dvipsnames]{xcolor}

\newcommand{\be}{\begin{equation}}
\newcommand{\ee}{\end{equation}}
\newcommand{\bea}{\begin{eqnarray}}
\newcommand{\beas}{\begin{eqnarray*}}
\newcommand{\eea}{\end{eqnarray}}
\newcommand{\eeas}{\end{eqnarray*}}
\newcommand{\ba}{\begin{array}}
\newcommand{\ea}{\end{array}}

\begin{document} 

\title{Are neutrino oscillation mixings linked to the smallness of solar neutrino scale?}

\author{Eduardo Becerra-Garc\'ia} \email{ebecerra@fis.cinvestav.mx}
\author{Abdel P\'erez-Lorenzana}
\email{aplorenz@fis.civestav.mx}

\affiliation{Departamento de F\'{\i}sica. Centro de Investigaci\'on y de Estudios Avanzados del I.P.N., \\ 
Apdo. Post. 14-740, 07000, Mexico City, Mexico.} 

\begin{abstract}
Observed reactor and atmospheric neutrino oscillation mixing values appear to be related to the neutrino scale ratio $\sqrt{\Delta m^2_{sol}/\Delta^2_{ATM}}$ in a way that suggest that the neutrino mass matrix can be expanded as a power series by using this ratio as the  smallness parameter. This approach provides a simple and natural way to expose the inner hierarchies among neutrino mass terms, which amounts to also explain the solar oscillation mixing as well as solar oscillation scale.
We explore a class of  mass matrix textures that realize this scenario, for both normal and inverted neutrino mass hierarchies, as well as CP violation and their stability under renormalization scaling.
\end{abstract}
\date{\today}
\maketitle

\section{Introduction}

The flavor problem, meaning, the understanding of the observed pattern of fermion masses and flavor mixings in the Standard Model (SM) of particles at a fundamental level, remains so far as an open question whose answer might shed light on possible new physics. This question is particularly more puzzling in the case of neutrinos which are predicted to be massless on the basis of the SM, and yet the abundant data collected along the last decades by several  oscillation neutrino experiments~\cite{nuoscPDG} undoubtedly indicates that neutrinos are rather light and  mix. Most of such data is well described with the three standard neutrinos. Global fits~\cite{nufit2021} give for the two independent mass scales
$\Delta m_{sol}^2 = 7.42^{+0.21}_{-0.20}\times 10^{-5}~eV^2$ and $\Delta m_{ATM}^2 = 2.510^{+0.027}_{-0.027} (2.490^{+0.026}_{-0.028})\times 10^{-3}~eV^2$,
corresponding to the ones that determine solar and atmospheric neutrino oscillation lengths, respectively, for the normal (inverted) hierarchy of masses.
As a remainder for the reader, hierarchy refers to the relative ordering of the mass eigenvalues, $m_i$ for $i= 1,2,3$,  such that if $m^2_3>m^2_1$ the mass pattern is said to be in a normal hierarchy (NH), otherwise it is known as an inverted hierarchy (IH).
Theoretically, $\Delta m_{sol}^2$ is identified as the squared mass difference $\Delta m_{21}^2$, where $\Delta m^2_{ij} = m_i^2-m_j^2$, whereas $\Delta m^2_{ATM}$ corresponds to $\Delta m^2_{31}$ ($|\Delta m^2_{32}|$), for NH (IH).   Note that the sign of  $\Delta m^2_{21}$ is already known due 
the MSW effect~\cite{MSW1,MSW2} that takes place within the sun, however the sign of $\Delta m^2_{31}$ and therefore the hierarchy of masses is still unknown. 

With massive neutrinos, mixings arise from the fact that neutrino mass terms are non diagonal in the flavor state basis which  differs from the mass eigenstate basis. Both the basis are connected by a unitary matrix $U_{mix}$  that in the Pontecorvo-Maki-Nakagawa-Sakata (PMNS)~\cite{pontecorvo,MNS} parameterization is written as $U_{mix}=VK$, where $K$ is a diagonal matrix containing two Majorana phases, and $V$ is given in terms of two real and a complex rotations, as  
$V = R_{23}(\theta_{23})R_{13}(\theta_{13},\delta_{CP})R_{12}(\theta_{12})$.
Same global fit analysis determines the three neutrino mixing angles to be
$\sin^2\theta_{12} = \sin^2\theta_{sol}=0.304^{+0.012(3)}_{-0.012}$, for the solar neutrino oscillations, $\sin^2\theta_{23}= \sin^2\theta_{ATM}=0.450^{+0.019}_{-0.016}(0.570^{+0.016}_{-0.022})$ for atmospheric ones, and $\sin^2\theta_{13}= 0.02246(1)^{+0.00062(74)}_{-0.00062}$, corresponding to reactor oscillation mixings. It is worth noticing  that the most recent fit from Ref.~\cite{nufit2021} changed drastically the best fit point on the atmospheric mixing for normal hierarchy, which lays now on the first octant, as compared to the previous fit in Ref.~\cite{nuglobal}, where it appeared on the second octant with the value $\sin^2\theta_{ATM}=0.573(5)^{+0.016}_{-0.020(19)}$. 
Yet, at three sigma level, on both fits the measured mixing regions still overlaps on basically the same area that extends on both the octants. Neutrino oscillations are also sensible to the Dirac CP phase which so far has been narrowed to be about $\delta_{CP}=230^{+36}_{-25}$ ($278^{+22}_{-30}$) and within the interval $[144^o,350^o]$ ($[194^o,345^o])$ at three sigma level. All above for NH (IH). Ongoing and next generation oscillation neutrino experiments will focus on increasing the precision on the determination of this  phase.

By looking up on these measured parameters it is unavoidable to notice some peculiarities that are very suggestive when they are put down together. First 
of all,  the hierarchy among solar and atmospheric scales are so that their ratio, in central fit values, can be expressed as 
\be \lambda= \sqrt{\frac{\Delta m^2_{sol}}{\Delta m^2_{ATM}}} =0.1719(26)~.
\label{lambda}
\ee
On the other hand, for the reactor  mixing one has
\be
\tan\theta_{13}= 0.1516\sim{\cal O}(\lambda)~.
\label{theta13vse}
\ee
Furthermore, the deviation of the atmospheric mixing from maximal turns out to be
\be
|\tan\theta_{ATM}-1| = 0.0955~(0.1513)\sim{\cal O}(\lambda/2)~[{\cal O}(\lambda)]~.
\label{thetaATMvse}
\ee
As stressed by above expressions, it is striking to note that all of those parameters are about the same order of magnitude. This rises the question of whether they could have a common source and exploring this possibility is the main goal of the present work. As it is clear, if the answer to such a question were on the positive,  the next issue would be to understand why solar mixing, being rather large but far from maximal, is not in evident connection with the same parameter. As we shall argue here, the solution to the whole riddle could indeed lay on the smallness of the solar scale, expressed through a simple neutrino mass matrix structure that emerges when one assumes that mass terms are functions that can be (Taylor) series expanded using $\lambda$ as the only smallness parameter. This comes out natural when one considers that physical masses are subjected to loop correction effects from  a yet unknown interaction characterized by a coupling parameter of order $\lambda$.
In such an approach, the inner hierarchy of the mass terms provides at the zero order
a maximal $\theta_{23}$ with a two degenerate neutrino spectrum, where other mixings are null. The order $\lambda$ corrections would  then be enough to predict all the mentioned mixing features, as well as to provide an understanding to solar mixing through the lifting of the degeneracy on the solar sector.

With this aim, this paper is organized as follows. First, 
we use the condition of having null reactor and solar  mixings and a maximal atmospheric mixing, with a two neutrino degenerate spectrum, to search for explicit mass matrix structures that be consistent with hierarchical neutrinos, based on the Taylor series approach. Then, along section three, we parametrize the order $\lambda$ contributions that rise the degeneracy, providing a  $\Delta m^2_{sol}$ mass gap,  which, at the same time, would offer an explanation for the observed neutrino mixings patterns. There we perform the analysis for both normal and inverted hierarchies. Henceforth,  in section four, we explore both analytically and numerically the possible predictions for CP violation of such scenarios, emphasizing that our approach can account for values within current observed bounds. The effect of running under renormalization group effects on the proposed mass matrix hierarchies is explored along section five. 
Finally, we close with some general concluding remarks, presented in section six.

\section{Linking neutrino mixings to solar scale} 
Based on the suggestive numerical coincidences depicted by Eqs.~(\ref{lambda}) to (\ref{thetaATMvse}), we propose that the scale ratio $\lambda$ can be used to Taylor expand the general neutrino mass matrix, in order to explore the hierarchies among the neutrino mass terms,  as dictated by  atmospheric and reactor mixings. Thus, we write the most general Majorana mass matrix as
\be
M_\nu = M_0 + \delta M(\lambda),
\label{Mtaylor}
\ee
where  $M_0$ stands for the zero order matrix of the expansion and $(\delta M)_{\ell\ell'}$, for $\ell,\ell' = e,\mu,\tau$, are all smooth complex functions of $\lambda$, at least of order one on it.  Of course, a leading higher power on $\lambda$ for some of these terms could  be possible, but that should be reflected along the calculations. Hence the above approach can be make without lost of  generality.

By taking the theoretical limit where $\Delta m^2_{12}$ is null, as well as the right hand side of expressions (\ref{theta13vse}) and (\ref{thetaATMvse}), one ends with a neutrino spectrum with two degenerate neutrinos where $\tan\theta_{23} = \pi/4$ with not further mixings. This scenario immediately suggest that, in the theoretical limit where $\lambda$ is null, the neutrino mass matrix has the simple form
\be
M_0 \propto \left(
\ba{ccc}
A & 0 & 0  \\
0 & 1 & \sigma C \\
0 & \sigma C & 1 \ea \right)~~,
\ee
where $\sigma$ stands for a positive (negative) sign that should be taken in the normal (inverted) hierarchy case. Here, $C$ is a complex number with  $Re\, C>0$,  
and $|A| = |1-\sigma C|$. In any case the only existing mass splitting at this limit is given by $\Delta m^2 \propto 4 Re\,C$. Also, $M_0$ does not generate any Dirac phase. Since the $A$ phase can be freely redefined, and taken to be zero, a single Majorana phase, $\phi$, arises from the phase difference of complex eigenmasses, $1\pm C$, which gives  $\tan\phi = 2Im\,C/(1 - |C|^2)$. 

Hereafter, to clearly exemplify the way our proposal works, we will concentrate our discussion to the case where $C = 1$, which means that we can take $|A| = (1-\sigma)$ according to the hierarchy. This would restrict our analysis to  truly hierarchical neutrinos.
Following  Eq.~(\ref{Mtaylor}),   the mass matrix should add to $M_0$  some smooth complex functions of $\lambda$, which by construction we assume at least  proportional to $\lambda$.  
In this terms, the Taylor form of the neutrino mass matrix 
can be written as
\be
M_\nu =  \left(
\ba{ccc}
\Gamma & e \lambda& a\lambda  \\
e \lambda & b & \sigma c \\
a\lambda & \sigma c & 1 \ea \right)~m_0~,
\label{Mnu}
\ee
where we naturally expect the real overall mass scale $m_0\sim \sqrt{\Delta m_{ATM}^2}/2$.
It is also worth noticing that in the $\mu$-$\tau$ block, from our construction principles,  generic mass terms $b$ and $c$ can be expressed as $1+\delta f \approx e^{i\varphi}(1+Re\,\delta f + |\delta f|^2)$, where $\delta f$ stands for the corresponding ${\cal O}(\lambda)$ complex correction. Hence, the overall phase turns out to be small, since $\tan\varphi\approx Im\,\delta f$. This already indicates that  the phases on the first row shall be the most relevant for the analysis below.
Furthermore, without loss of generality, upon rephasing of the flavor neutrino wave functions, and without altering the   phases relevance, 
we can choose $e$ and $b$  to be real parameters. Others should, in general, stay complex numbers.

According to our Taylor expansion considerations,
 at the leading order, we can write
$|\Gamma| = (1-
\sigma)+ d\lambda$, $b = 1+\sigma \delta_b\lambda$, and $|c|=  1 - \delta_c \lambda $. 
Here we should assume $|a|$, $d$, $e$, $\delta_b$, and $\delta_c$ to be at most order one numerical parameters to be fixed by the use of experimental data. Also, hereafter we denote the phases of $\Gamma$, $a$ and $c$ as $\phi_{\Gamma,a,c}$, respectively, where, so far, $\phi_c$ should be assumed to be an small one.

A similar texture to the above, for normal hierarchy, was presented in Ref.~\cite{solmt}, motivated within the context of $\mu-\tau$ symmetry. There, however, the corresponding corrections to $m_{ee}$ and $m_{\mu\tau}$ where proposed to be of order $\lambda^2$, but no connection with our current motivation was made. 
Such a case, of course, is somewhat contained in the present one since our approach is more general.
Note also that a class of models based on $\mu-\tau$ symmetry or $\mu-\tau$ parity can also be reproduced in the present context (for a recent review of both see for instance~\cite{mtreview}). That should be expected, since such symmetries are known to be closely attached to the smallness of $\theta_{13}$ and a large $\theta_{23}$.  Furthermore, Eq.~(\ref{Mnu}) already shows some of the inner hierarchies that the observed neutrino oscillation data suggests.

\section{Masses and mixing} 
Next, let us proceed to show that the above expression for $M_\nu$ does predict the desired spectra of masses and mixing. It would be of particular interest to verify that the right amount of solar mixing could arise from the texture. Because at the zero order $m_{1,2}^2$ are degenerated, one only needs the right amount of perturbation to split these squared masses to provide a  mass gap $\Delta m^2_{21}\propto \lambda^2m_0^2$. As for the large solar mixing,  it should naturally arise due to the same kind of perturbations. 

In order to estimate the mixings, we perform an approximate diagonalization  of the squared hermitian matrix $H= M_\nu M_\nu^\dagger$.
This has the clear advantage that $H$  
is diagonalized by a unitary transformation, such that $U^\dagger HU = M_{diag}^2$, where the RHS is given by the non negative and diagonal matrix $M_{diag}^2 = diag(m_1^2,m_2^2,m_3^2)$. 
Up to field phase redefinitions of the neutrino basis, $U^*$ corresponds to the $U_{PMNS}$ mixing matrix that diagonalizes $M_\nu$ (see Refs.~\cite{takagi,horn} for the precedents to this approach).
For the purpose of calculation, it becomes convenient, given the expected smallness of $\theta_{13}$ and the hierarchical nature of neutrinos, to express $U =R_{23}(\theta_{23},\varphi_1)R_{13}(\theta_{13},\varphi_2)R_{12}(\theta_{12},\varphi_3)$,  where the block unitary rotations 
\be \label{R}
R(\theta,\varphi) = \left(
\ba{cc}
\cos\theta & \sin\theta e^{i\varphi} \\
-\sin\theta e^{-i\varphi}& \cos\theta
\ea
\right)~,
\ee
embedded into their three dimensional forms,  are obtained by consecutively diagonalizing each indicated block on $H$. On this approach, Dirac CP phase on PMNS parameterization turns out to be $\delta_{CP} = \varphi_2 -\varphi_1-\varphi_3$. 

Therefore, within our present setup, $H$ can be written as
\be
H =  \left(
\ba{ccc}
\alpha & \delta& \omega  \\
\delta^* & \beta & \rho \\
\omega^* & \rho^* & \gamma \ea \right)~m_0^2~,
\label{H}
\ee
where,  
$\alpha = |\Gamma|^2 + e^2\lambda^2 + |a|^2\lambda^2$, 
$\beta =  b^2 + |c|^2 + e^2\lambda^2$, 
$\gamma = 1+ |c|^2+|a|^2\lambda^2$, 
$\delta = (\Gamma e + eb  +\sigma ac^*)\lambda$, 
$\omega = (\Gamma a^* + \sigma ec^* + a)\lambda$, and 
$\rho = \sigma bc^* + \sigma c +ea^*\lambda^2$. 
It is then easy to realize that in normal hierarchy $\alpha$ becomes ${\cal O}(\lambda^2)$, as well as the perturbative corrections that would be induced by off-diagonal matrix terms, $\delta$ and $\omega$. This would explain solar neutrino scale without further assumptions. Such is, however, not the case for inverted hierarchy, where $\alpha$ gets an order one correction on $\lambda$. Here, explaining a solar scale that is of second order would require tuning some of the parameters to an order $\lambda$ precision,  as we will discuss below. 

After some algebra, atmospheric mixing is expressed, at leading order, as
\be
\tan\theta_{ATM} \approx 1 +\frac{\delta_b}{2|\cos\phi_c|} \lambda~,
\label{tanATM}
\ee
irrespective of the hierarchy, 
whereas the reactor mixing becomes
\be
\tan\theta_{13} \approx \frac{|(1-\sigma) e^{i\phi_a} (e+a^*) + (e+a)(1+\sigma e^{-i\phi_c}) |}{\sqrt{8}(1+|\cos\phi_c|)}~\lambda~.
\label{tan13}
\ee
In order to fulfill the expectation of having at most order one Taylor parameters, $|\cos\phi_c|$ should be close to one,  otherwise the coefficient of $\lambda$ on Eq.~(\ref{tanATM}) would become too large as to account for the observable value of the mixing. Moreover, since reactor mixing in normal hierarchy vanishes for $\phi_c = \pi$, we must avoid this branch of values. This is of course consistent with our previous expectations. In the particular case where we take $\phi_c\approx 0$, that we assume hereafter, atmospheric and reactor mixings reduce to
\be
\tan\theta_{ATM} \approx 1 +\frac{\delta_b}{2}\lambda~,\quad \text{and}\quad
\tan\theta_{13} = \frac{|e+a|}{2\sqrt{2}}~\lambda~, 
\label{tans}
\ee
for both the hierarchies. Furthermore, by looking up on best fit value of the atmospheric  mixing, we should notice that $\delta_b$ has to be negative for NH, in order to bring down the mixing from its zero order maximal value. The opposite should be the case for IH.
From here it is straightforward to see that, at best fit point, $\delta_b = -1.11(1.753)$ and $|e+a| = 2.494(81)$ according to our initial assumptions.
As an aside comment, it is obvious that above formula for atmospheric mixing would still work pretty well  even if the resulting best fit point from future data were to move towards values higher than $\pi/4$, as it was in the data fit of Ref.~\cite{nuglobal}, for instance. 

Next, we proceed to calculate  the solar neutrino scale and the solar mixing, which after a lengthy calculation get the general forms
\be
\Delta m_{sol}^2\approx m_0^2 \sqrt{\mu^2 + D^2}~,
\label{msolar}
\ee
and 
\be
\tan\theta_{sol} \approx \frac{1}{x +\sqrt{1+x^2}}~,
\label{tansol}
\ee
with $x=\mu/D$ and 
where the explicit expressions for $D$ and $\mu$ do depend on neutrino hierarchy. From Eq.~(\ref{msolar}), it is clear that both $\mu$ and $D$ should be required to be $\sim {\cal O}(\lambda^2)$, which in turn would imply that $0<x\alt {\cal O}(1)$, just as needed to understand the large observed value of solar mixing. As a matter of fact, at best fit point data requires that $x = 0.426$. The corresponding results to each hierarchy are as follows.

(i)  For {\it normal hierarchy} we get at the lower order
\be
D \approx \frac{1}{\sqrt{2}}\left|(e-a)(\delta_b +2\delta_c)+ 2d e^{i\phi_\Gamma}(e -a^*)\right|\lambda^2~,
\ee
whereas, 
\be
\mu \approx \frac{1}{4}\left((\delta_b+2\delta_c)^2-4d^2\right)\lambda^2~.
\ee

(ii) For {\it inverted hierarchy} the situation is more challenging, since, as we have anticipated, both contributions to the solar scale, $\mu$ and $D$, arise at first order level, such that
\be
D \approx 2\sqrt{2}\left|e(1+e^{i\phi_\Gamma}) - (a+a^*e^{i\phi_\Gamma})\right|\lambda +{\cal O}(\lambda^2)~,
\ee
and
\be
\mu \approx -\left(4d +2(\delta_b+2\delta_c)\right)\lambda+{\cal O}(\lambda^2)~.
\ee
Thus, in order to warrant the right prediction for the solar scale, some tuning conditions must be imposed on the parameters, such that they cancel each other within the order $\lambda$ contributions. From above expressions, it is straightforward to see that such conditions amount to fix  
\be
a \approx e- g\lambda~,\qquad\text{and}\qquad d \approx -\frac{1}{2}(\delta_b + 2\delta_c) - h\lambda~,
\label{tuning}
\ee
with $|g|$ and $h$ order one numbers. This express the required level of accuracy of our approximation. It has the aside implication that $\phi_a$ would become small (and perhaps irrelevant for CP violation), since now $\tan\phi_a\approx -Im(g/e)\cdot\lambda$~. Notice that no restrictions are made to $\phi_\Gamma$ so far, which should then become the most relevant CP phase in this case. 

Using the above tuning conditions, we then get
\be
D\approx \sqrt{2}\left| e\delta_b(1+e^{i\phi_\Gamma})-2(g+g^*e^{i\phi_\Gamma}) \right|\lambda^2,
\ee
and 
\be
\mu \approx \left(\frac{1}{2}\delta_b^2 + 2h -|e+a|^2\right)\lambda^2~.
\ee

Finally, atmospheric scale is given, for both the hierarchies as
\be
\Delta m^2_{ATM}\approx \left(4 +(\sigma\delta_b-2\delta_c)\lambda\right)m_0^2~.
\label{matm}
\ee

Since $\delta_b$ has been fixed by the observed value of the atmospheric mixing, in Eq.~(\ref{tanATM}), the model is left with five free parameters: That is, $m_0$, $e$, $\delta_c$ and $a$ and $d$  for normal hierarchy, but $g$ and $h$ for the inverted one. Hence, there is plenty of room to accommodate the remaining four neutrino observables, $\tan\theta_{13}$, $\tan\theta_{sol}$, $\Delta m^2_{sol}$ and $\Delta m^2_{ATM}$, from Eqs.~(\ref{tans}), (\ref{msolar}),  (\ref{tansol}) and (\ref{matm}).
To have a feeling of the order of magnitude the model parameters should have,  we can solve the system in the non CP violating case, by fixing $m_0 = \sqrt{\Delta m^2_{ATM}}/2 =0.025~eV$, which  gives at central values $a = 0.664$, $d = 0.61$, $e=1.829$ and $\delta_c = 1.867$ for normal hierarchy, and 
$e = 1.24$, $g = 0.437 $, $h=3.092$ and $\delta_c = 0.877$ 
for 
the inverted one. Of course, taking a slightly different value for $m_0$ would render a bit different set of solutions, but all would be of similar order, which validates our approximations. Note, also,  that these values emerge for the approximated expressions we have derived above. More precise results can be obtained either from extending our analysis to include higher order corrections or by proceeding through a completely numerical calculation. We shall consider the later below. 

\section{CP violation}
The amount of CP violation in any neutrino mass model can be estimated by using the Jarlskog invariant~\cite{jarlskog}
\be
J =\frac{1}{8}\sin2\theta_{12}\sin2\theta_{23}\sin2\theta_{13}\cos\theta_{13}\sin\delta_{CP}~,
\ee
which can also be written in terms of the previously defined squared hermitian matrix $H$ as (See for instance~\cite{branco})
\be
J = -
\frac{Im\left(H_{21}H_{32}H_{13}\right)}{\Delta m^2_{21}\Delta m^2_{31}\Delta m^2_{32}}~.
\ee
Using our parameterization, we can straightforwardly express the invariant in the approximated form
\be
J\approx -\left(\frac{m_0^2}{\Delta m^2_{ATM}}\right)^3~Im[\delta^*\rho^*\omega]\,\lambda^{-2}.
\ee
It turns out that, at the leading order, $Im[\delta^*\rho^*\omega]\sim{\cal}(\lambda^3)$, which means that  the model could without stress accommodate a $\sin\delta_{CP}\sim {\cal O}(\lambda)$ which is well within the expected experimental value range. As a matter of fact,  
by writing down
\be 
Im[\delta^*\rho^*\omega]\,\lambda^{-2}\approx \kappa_0 +\kappa_1\lambda + \dots ~,
\ee
and after some calculations, we get, for normal hierarchy,  
$\kappa_0 = (|a|^2 -e^2)\sin2\phi_c$ and 
\begin{eqnarray}
\kappa_1 &=& \left[\delta_b(|a|^2 - e^2)+2e^2\delta_c\right]\sin2\phi_c+2\cos\phi_c\Big[e|a|(\delta_b + 2\delta_c)\sin\phi_a +  \nonumber \\ &&
e^2(\delta_c-d)\sin(\phi_\Gamma+\phi_c)+|a|^2d\sin(\phi_\Gamma+\phi_c-2\phi_a)\Big]~.
\eea
Notice that this implies that $\kappa_0\alt{\cal O}(\lambda)$ as a consequence of the smallness of $\phi_c$, just as discussed along previous sections. Furthermore,  in the limit where we do neglect $\phi_c$,  $\kappa_0 = 0$, and thus,
the CP violation phase is given by 
\be
\sin\delta_{CP}\approx -0.466~ \left(\frac{4m_0^2}{\Delta m^2_{ATM}}\right)^3\, \kappa_1\,\lambda~,
\ee
where $\kappa_1\approx 2[e|a|(\delta_b + 2\delta_c)\sin\phi_a +  e^2(\delta_c-d)\sin\phi_\Gamma+|a|^2d\sin(\phi_\Gamma-2\phi_a)]$~. It is not difficult to see that with order one parameters we can easily get an adequate value for the CP violating phase.  

For inverted hierarchy we get
$$
\kappa_0 = 2\cos\phi_c\Big[2\left(|a|^2\sin(\phi_\Gamma+\phi_c-2\phi_a)-e^2\sin(\phi_\Gamma+\phi_c)\right) +(|a|^2-e^2)\sin\phi_c +4e|a|\sin\phi_a\Big]~,
$$
and the rather complicated expression
\begin{eqnarray}
\kappa_1 &=& 2\delta_b\Big[2e|a|\sin(\phi_c-\phi_a) - |a|^2\sin(2\phi_c+\phi_\Gamma-2\phi_a)+e^2\sin\phi_\Gamma\Big]  \nonumber \\
&&
+ ( e^2(2\delta_c - \delta_b) - \delta_b|a|^2 )\sin2\phi_c
-2\cos\phi_c\Big[e|a|(6\delta_c - \delta_b-4d)\sin\phi_a 
 \\
&&
 + 2e|a|\delta_b\sin(\phi_a-\phi_\Gamma) + e^2(d - \delta_c)\sin(\phi_\Gamma+\phi_c)
- |a|^2(d-4\delta_c)\sin(\phi_\Gamma +\phi_c -2\phi_a)\Big].
\nonumber
\end{eqnarray}
In this case, the conditions imposed through Eq.~(\ref{tuning}), and the implied smallness of $\phi_{c,a}$ amount again to consider $\kappa_0$ to be of order $\lambda$. 
Indeed, by neglecting these phases in above expressions, they simplify into
$\kappa_0 \approx  - 8eRe(g)\sin\phi_\Gamma\,\lambda$, and $\kappa_1 =e^2(4\delta_b+  6\delta_c) \sin\phi_\Gamma $~. Hence, for inverted hierarchy we get the approximated formula
\be
\sin\delta_{CP}\approx -0.4688~ \left(\frac{4m_0^2}{\Delta m^2_{ATM}}\right)^3\, (\kappa_0 +\kappa_1\,\lambda)~,
\ee
which is now controlled by the single $\phi_\Gamma$ phase, as already anticipated.

In order to explore in a more accurate way the possible predictions for the Dirac CP phase from our  neutrino mass matrix proposal, we have numerically scanned the parameter space looking for value sets  that reproduce the observed mixings and oscillation scales within the current level of accuracy. To this aim, 
we numerically solve for the eigensystem of  the hermitian matrix $H$ that is generated out of the mass matrix $M_\nu$, in Eq.~(\ref{Mnu}), by randomly assigning values to the parameters. Once the eigenvectors are numerically known,  without further considerations nor approximations, we calculate the corresponding mixing angles and the Dirac CP phase as given in the PMNS parameterization, as well as the corresponding squared mass differences, and keep only those parameter sets that reproduce neutrino oscillation observables within the range of one sigma deviation for the mixings, and up to three sigma deviations for the mass scales ratio. Given that $m_0$ appears in Eq.~(\ref{Mnu}) as a global scaling, we have found more appropriate to use the scale ratio as the discriminator rather than the scales themselves since this reduces one parameter on the analysis.

Also, in order to simplify the numerical computation
we fixed our small parameter to the value 
$\lambda = 0.1719$ throughout the numerical analysis for both hierarchies, with the understanding that any possible change on it can always be compensated by a scaling on the other parameters. It is also worth noticing that for any given matrix $H$ which is diagonalized by the unitary matrix $U$, its complex conjugated $H^*$ would be diagonalized by $U^*$. Since the Dirac phase in $U$ does correspond to the opposite phase in $U^*$, the parameter map for $\delta_{CP}$ should be symmetric under reflection symmetry. We use this fact to locate additional point solutions on the parameter space.

\begin{figure}[ht]
  \centering
    \includegraphics[width=0.8\linewidth]{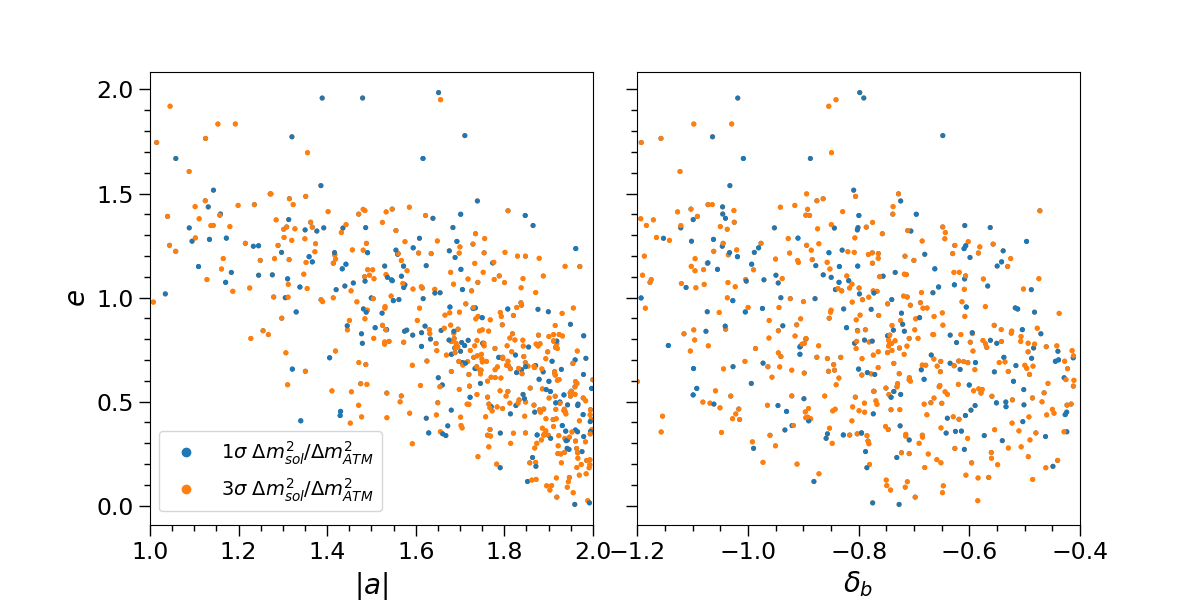}
    \caption{Allowed parameter space for the normal hierarchy, $e$ is shown as a function of $|a|$ and $\delta_b$. Blue (orange) dots generate a  $\Delta m_{sol}^2/\Delta m_{ATM}^2$ ratio at $1\sigma~(3\sigma)$ accuracy.}
    \label{e_a_deltab}
\end{figure}

For normal hierarchy we searched the parameter space using the following parameter ranges: $|a|\in(1,2)$, $d\in(0,2)$, $e\in(0,2)$, $\delta_b \in (-1.2,-0.4)$, $\delta_c \in (0,2)$ and $\phi_{a,c,\Gamma} \in (0,2\pi)$. 
We show in Fig.~\ref{e_a_deltab} the solution sets for the $e$-$|a|$ and $e$-$\delta_b$ subspaces, where the allowed region is best depicted. Notice the apparent correlation among $e$ and $|a|$, that was expected from the way they combine themselves to produce the reactor mixing in Eq.~(\ref{tan13}).
Also, all the presented solutions were found to have a very small value of $\phi_c$, as expected from our previous analytical discussions. 
Blue  dots correspond to the values of the parameters which generate a $\Delta m_{sol}^2/\Delta m_{ATM}^2$ ratio within the $1\sigma$ range, while orange dots correspond to those solutions with the quotient lying in the $3\sigma$ range. 
The same colormap is used for all the figures presented in this work. As we can see from the plots, there is not a clear distinction between the $1\sigma$ and $3\sigma$ regions, indicating the sensibility of the mass texture to the values of the parameters regarding the prediction of neutrino mass scales.
This may be an indication that certain level of tuning would be required in the underlying theory to keep masses under the expected accuracy.

For the inverted hierarchy, we have explicitly introduced the tuning conditions given in Eq.~(\ref{tuning}) to focus our search. Thus, the parameters were assigned to arbitrary values within the intervals: $|g| \in (0,2)$, $h \in (1,2)$, $e \in (1,1.5)$, $\delta_b \in (1,2)$, $\delta_c \in (0,1.8)$ and $\phi_{c,g,\Gamma} \in (0,2\pi)$.
Fig.~\ref{e_h_g} shows the results on the $e$, $|g|$ and $h$ subspace. Unlike the parameter space for the NH case, which present just a slight correlation only between two of them, for the inverted hierarchy there is a clear correlation between all parameters. For the sake of simplicity we only show here three plots, however, $\delta_b$ and $\delta_c$ have shown a similar dependence in $|g|$ as that for $e$ and $h$. 

\begin{figure}[ht]
  \centering
    \includegraphics[width=0.6\linewidth]{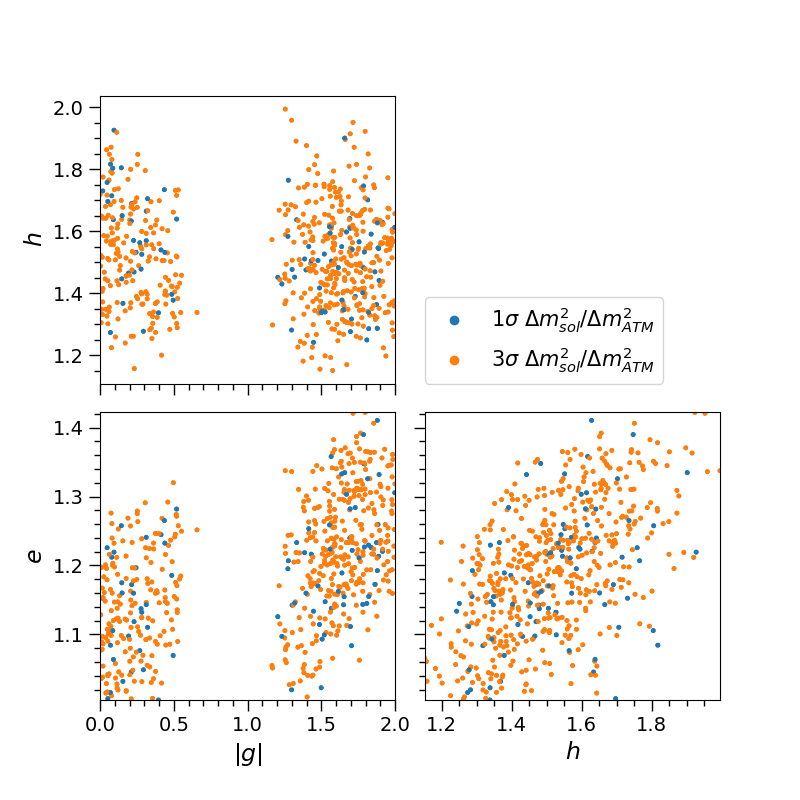}
    \caption{Allowed parameter values for the inverted hierarchy as seen in the $e$-$|g|$-$h$  parameter subspace.}
    \label{e_h_g}
\end{figure}

In Fig.~\ref{phigamma_g} we show the phase $\phi_\Gamma$ as a function of $|g|$. This phase is of particular interest because
our results were also found to allow only small values for the $\phi_{c,g}$ phases, as expected.  Therefore, it is the $\phi_\Gamma$ phase the one that would 
determine the overall shape of the Dirac CP phase. Furthermore, as we can observe on Figs.~\ref{e_h_g} and \ref{phigamma_g}, there are well delimited regions in the parameter space, although, once again, the points which generate a mass scale ratio within $1\sigma$ and $3\sigma$ ranges are mixed together. Thus, the already observed sensibility of the mass scales in NH to small variations of the parameters also appears for IH.

\begin{figure}[ht]
  \centering
    \includegraphics[width=0.5\linewidth]{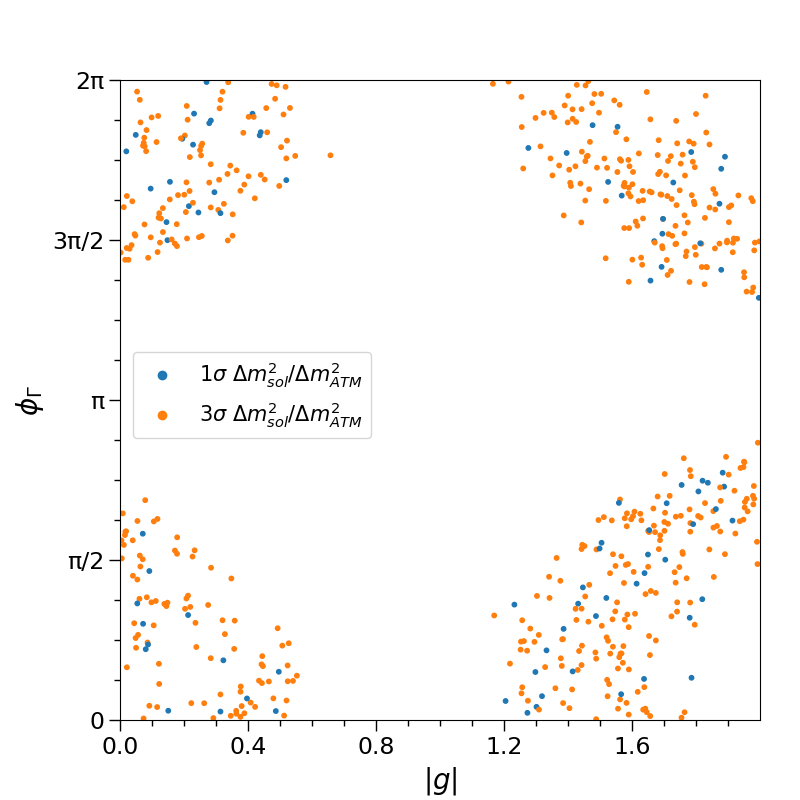}
    \caption{$\phi_\Gamma$ as a function of $|g|$. There are well delimited regions for the values of the model parameters which generate neutrino oscillation parameters within the required accuracy.}
    \label{phigamma_g}
\end{figure}

The predicted values of the Dirac phase that emerge from the found parameter sets that are consistent with oscillation  mixings and mass scales at the given level of accuracy,
are depicted in Figs.~\ref{deltaCP_NH} and \ref{deltaCP_IH} for normal and inverted hierarchy, respectively. The blue (orange) band indicates the current $1\sigma$ ($3\sigma$) region for $\delta_{CP}$, as reported in Ref.~\cite{nufit2021}. 
As we can observe, the model at hand has quite enough room on the parameter space to allocate a future measurement of $\delta_{CP}$ well within the current expectations. 
Notice that
in the NH case the majority of the points lying within the $1\sigma$ range for $\delta_{CP}$ correspond approximately to $e \in (0.2, 0.8)$, whereas in the IH case, we have fewer points within the $1\sigma$ and $3\sigma$ regions, these lying approximately in the interval $g \in (0,0.5)\cup (1.5,2)$. Thus, indeed, the  numerical analysis shows that there exist a parameter space for the proposed mass texture consistent with the $1\sigma$ values of masses and mixings in both cases. This validates the interest on our phenomenological approximation and enforces the possible link among the observables that suggested it.
 
\begin{figure}[ht]
  \centering
    \includegraphics[width=0.5\linewidth]{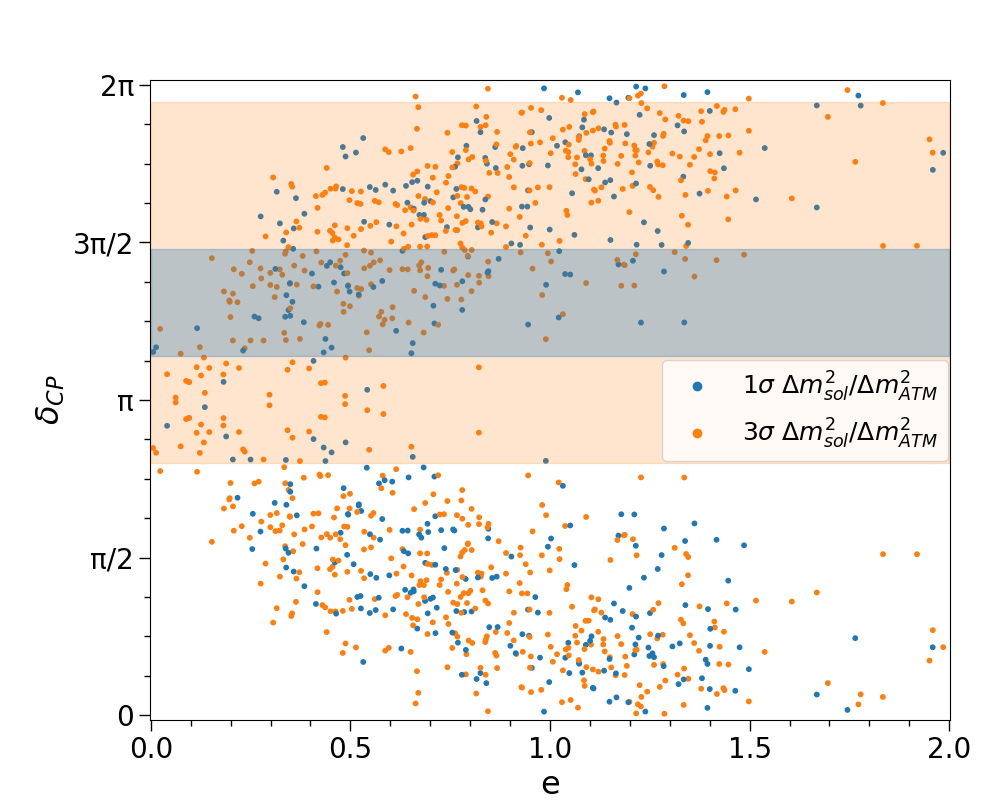}
    \caption{$\delta_{CP}$ predictions in terms of the $e$ parameter for the normal hierarchy. The current allowed values are within the colored bands, as explained in the text.}
    \label{deltaCP_NH}
\end{figure}

\begin{figure}[ht]
  \centering
    \includegraphics[width=0.5\linewidth]{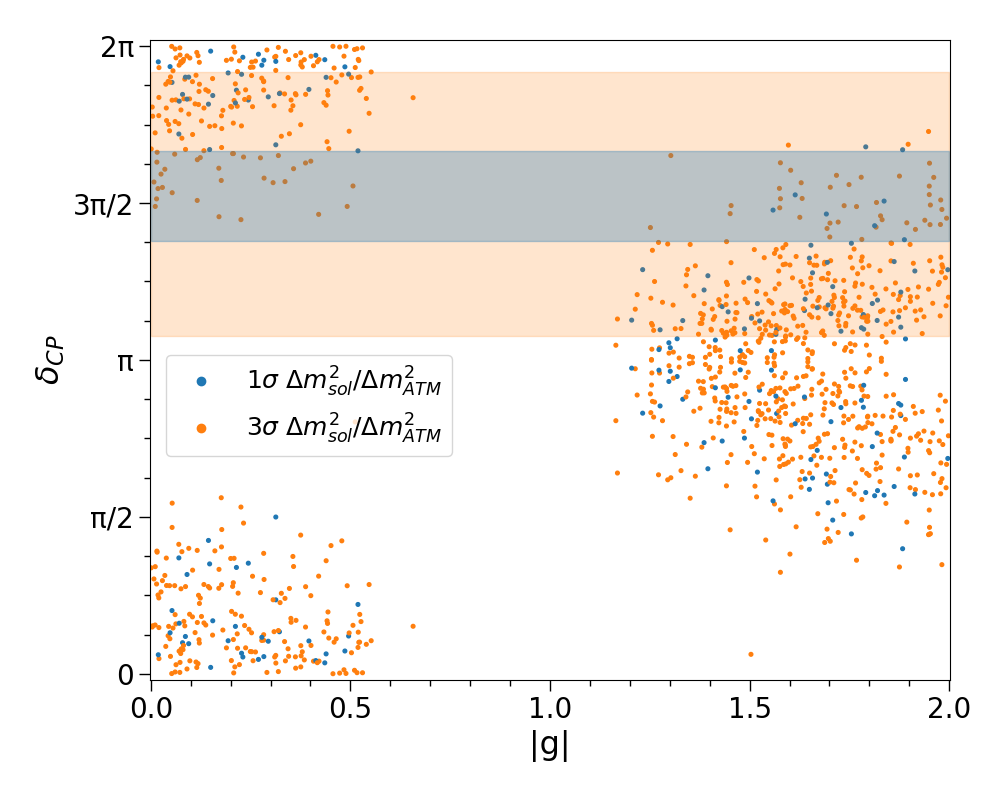}
    \caption{$\delta_{CP}$ predictions in terms of the $|g|$ parameter for the inverted hierarchy.}
    \label{deltaCP_IH}
\end{figure}

\section{Renormalization effects}
As a final note we would like to comment on the effects that energy scaling would have over the mass matrix structure that has been derived in above, by using the Taylor expansion approximation. This issue  arises as a necessary analysis when one considers that any realistic flavor model of neutrino masses should  
rely on some new physics, either in the form of new interactions or new particles, that very likely would emerge at some high energy $\Lambda$. Therefore, to reproduce what we measure at a low energy, $\Lambda_{0}$, below electroweak scale, the effect of running  parameters of the theory from $\Lambda$ down to $\Lambda_{0}$ has to be considered. This running is governed by the renormalization group equations (RGE), which, at one-loop, relate the effective neutrino mass matrix with that defined at higher energy by~\cite{RGE,RGE2,RGE3}
\be
M_\nu(\Lambda_0) = I_\zeta~ {\cal I}M_\nu(\Lambda) {\cal I}~,
\ee
where we assume that $M(\Lambda)$ has the structure  given in Eq.~(\ref{Mnu}).  $I_\zeta$ is the 
scaling factor 
\be
I_\zeta = exp\left[-\frac{1}{16\pi^2}\int_{\ln\Lambda_0}^{\ln\Lambda}\zeta(t)~dt\right]~,
\ee
which only contributes to  rescale the overall neutrino scale, but has no relevance on the reormalization of the mixing angles. Here, 
with the Standard Model content, $\zeta =-3g_2^2 + \lambda_H + 6 y_t^2 $, where  $g_2$ stands for the $SU(2)_L$ gauge coupling, $\lambda_H$ for the Higgs self coupling and  $y_{t}$  for the top Yukawa coupling.
On the other hand, the flavour dependent components of the diagonal matrix ${\cal I} = Diag\{I_e,I_\mu,I_\tau \}$, are given by 
\be
I_\ell = exp\left[\frac{3}{32\pi^2}\int_{\ln\Lambda_0}^{\ln\Lambda}y_\ell^2~dt\right]~,
\ee
with   $y_{\ell}$  the lepton Yukawa couplings. Due to charged lepton hierarchy, from which $y_e<<y_\mu<<y_\tau<<1$, the scaling factors $I_\ell$ would be close to unity and $I_\tau$ shall be the dominant one. Because of this, we can safely assume that $I_{e,\mu}\approx 1$ whereas $I_\tau=e^{r_\tau}\approx 1 + r_\tau$, where
\be
r_\tau \approx \frac{3}{32\pi^2}y^2_\tau(\Lambda_0)\ln\left(\frac{\Lambda_0}{\Lambda}\right)~.
\ee
As a matter of fact, by assuming  $\Lambda = \Lambda_{GUT}\sim 10^{16}~GeV$ and $\Lambda_0$ the electroweak scale, one gets that $r_\tau\sim 10^{-5}$, which represent a too small correction when compared to  $\lambda$, which defines the inner hierarchy of the mass matrix --actually $r_\tau\sim {\cal O}(\lambda^6)$--. Therefore, we would not expect RGE scaling effects to substantially alter the inner structure of the mass matrix. We can verify this by explicitly writing 
\be
M(\Lambda_0) = 
\left(
\ba{ccc}
\Gamma_R & e_R \lambda& a_R\lambda  \\
e_R \lambda & b_R & \sigma c_R \\
a_R\lambda & \sigma c_R & 1 \ea \right)~m_{0R}(\Lambda_0)~,
\ee
where the renormalized parameters stand for $\Gamma_R \approx \Gamma (1-2r_\tau)$, $a_R \approx a(1-r_\tau)$, $b_R\approx b(1-2r_\tau)$, $c_R\approx c(1-r_\tau)$, $e_R\approx e (1-2r_\tau)$ and $m_{0R}(\Lambda_0) = m_0 I_\zeta I_\tau^2$. From here it is clear that the whole effect of the RGE scaling, from SM corrections, would be rather negligible as far as the size of the parameters goes. Nevertheless, it is worth noticing that the relative scaling on $c_R$ respect to the ones on $b_R$ and $\Gamma_R$ would produce a slight splitting among $m_{1,2}^2$  that still will be much smaller than the required by solar neutrino scale. This situation is particularly most prominent for the inverted hierarchy, although still remains under control.

\section{Concluding remarks and outlook}
 
Measured reactor mixing and the deviation of the atmospheric mixing from maximal value, as given in Eqs.~(\ref{theta13vse}) and (\ref{thetaATMvse}) respectively, are about the same order of magnitude of the squared root of the ratio among neutrino oscillation scales. This curious observation could be more than just a numerical coincidence and rather be an indication of a common physical origin of those parameters. Here, we  have elaborated over the idea that all these value parameters may arise from perturbative corrections, to the neutrino mass matrix, that depend on powers of a single small parameter, $\lambda$, such that  mass terms can be expressed as power series on the last. Clearly, this idea would be well motivated, as well as natural, if there were hidden interactions in the neutrino sector, where loop corrections would be the ones responsible of fixing the (bare) zero order mass structure.  Although we do not know yet what such interactions might be, the simplest scenario is to consider at least a single coupling parameter, which could be interpreted as $\lambda$. 
 
In such an approximation, we have shown that the class of neutrino textures that arise at the zero order, for either normal or inverted hierarchies,  are quite unique. They are required to comply with having two degenerate neutrino states and null $\theta_{13}$ and $\theta_{12}$ but  a maximal $\theta_{23}$ mixing angles, which fixes the initial mass structure. Furthermore, the addition of linear corrections in $\lambda$ are quite enough to  generate the masses and mixings observed from neutrino oscillation experiments, using only factor coefficients that at most are of order one. The predicted CP phase is also compatible with the current experimental bounds at $1\sigma$ and $3\sigma$ accuracy. And, as we have shown, the proposed mass structures are stable under the energy scaling goberned by RGE with the SM particle content.
Our results strongly support the validity of approaching mass terms with the suggested power series approximation, in particular  to describe the inner hierarchies of the mass matrix. It would be interesting to explore this idea even further. 

The present modeling of the neutrino mass structure, however, seems to require some level of tuning on the parameters in order to provide precise predictions on neutrino  mass scales,  as our numerical exploration suggests. The required tuning should be of order $\lambda^2$ or so on most parameters, though. Mixing predictions, on the other hand, are quite robust. This feature of the model may suggest the existence of further correlations among the different mass terms, and perhaps be indicative of a non trivial interplay among the parameters of the possible underlying theory. It would be interesting to explore this  possibility in an extended way once a complete theoretical realization of the present idea could be provided. 

The texture suggested from our series expansion proposal may be obtained in models with $\mu-\tau$ symmetry, or $\mu-\tau$ parity, although we have not explored this possibility any further. The zero order mass matrix in the case of normal hierarchy seems consistent with models with a $L_e$ symmetry.   This may suggest the kind of flavor symmetries to further explore in search for a complete model realization of our proposal. In particular, models with $S_3$, $A_4$ or other discrete flavor symmetries which may hold a doublet fermion representation could explain the structure in the $2-3$ sector, which has almost universal mass terms before $\lambda$ perturbations are introduced.

\acknowledgments
Work partially supported by Conacyt, Mexico, under FORDECYT-PRONACES grant No. 490769. The work of E.~Becerra-Garc\'ia was supported by a CONACyT graduate fellowship.


\begin{thebibliography}{99}

\bibitem{nuoscPDG}
For a review see: M.C. Gonzalez-Garcia and  M. Yokoyama in P.A. Zyla et al. (Particle Data Group), Prog. Theor. Exp. Phys. {\bf 2020}, 083C01 (2020) and references therein.

\bibitem{nufit2021}
M.C. Gonzalez-Garcia, M. Maltoni, T. Schwetz,
Universe {\bf 2021}, 7(12), 459 [arXiv:2111.03086].

\bibitem{MSW1}
L. Wolfenstein, Phys. Rev. D {\bf 17}, 2369 (1978); Phys.  Rev.  D {\bf 20}, 2634 (1979). 

\bibitem{MSW2}
S.P. Mikheyev and A. Yu. Smirnov, Yad. Fiz. {\bf 42},  1441 (1985).

\bibitem{pontecorvo}
B. Pontecorvo, J. Exptl. Theoret. Phys. {\bf 33}, 549 (1957) [Sov. Phys.
JETP \textbf{6}, 429 (1958)]; J. Exptl. Theoret. Phys. {\bf 34}, 247 (1958)
[Sov. Phys. JETP \textbf{7}, 172 (1958)].  

\bibitem{MNS}
Z. Maki, M. Nakagawa, and S. Sakata, Prog. Theor. Phys. \textbf{28}, 870 (1962).

\bibitem{nuglobal}
I. Esteban, M. C. Gonzalez-Garcia,  M. Maltoni, T. Schwetz, A. Zhoue,  J. High Energ. Phys. 2020, 178 (2020) [arXiv:2007.14792].

\bibitem{solmt} 
A. P\'erez-Lorenzana, arXiv:2012.12243.


\bibitem{mtreview}
Z. Xing and  Z. Zhao, Rept. Prog. Phys. {\bf 79}, 076201 (2016). 

\bibitem{takagi}
T. Takagi. Japan. J. Math. {\bf 1}, 83-93 (1925).

\bibitem{horn}
R.A. Horn and  C.R. Johnson, {\it Matrix Analysis}, Cambridge University Press, New York, NY (1985).


\bibitem{jarlskog}
C. Jarlskog, Phys. Rev. Lett. {\bf 55}, 1039 (1985).

\bibitem{branco}
G.C. Branco, R. Conzalez Felipe, F. R. Joaquim, I. Masina, M.N. Rebelo and C. A. Savoy, Phys. Rev. {\bf D67}, 073025 (2003).



\bibitem{RGE}
P. H. Chankowski and Z. Pluciennik, Phys. Lett. B{\bf 316}, 312 (1993).

\bibitem{RGE2}
K. S. Babu,
C. N. Leung and J. T. Pantaleone, Phys. Lett. B{\bf 319}, 191 (1993).

\bibitem{RGE3}
S. Antusch,
M. Drees, J. Kersten, M. Lindner and M. Ratz, Phys. Lett. B{\bf 519}, 238 (2001).



\end{thebibliography}
\end{document}